\documentstyle[12pt]{article}

\font\mybb=msbm10 at 11pt
\def\bb#1{\hbox{\mybb#1}}
\def\ZZ{\bb{Z}}
\def\RR{\bb{R}}

\begin{document}
\thispagestyle{empty}
{\baselineskip=12pt
\hfill CALT-68-2034

\hfill hep-th/9601077

\vspace{1.0cm}}
\centerline{\large \bf M Theory Extensions of 
T Duality\footnote{Work supported in part by the U.S. Dept. of Energy
under Grant No. DE-FG03-92-ER40701.}}
\bigskip
\bigskip
\centerline{\large John H. Schwarz\footnote{Email: jhs@theory.caltech.edu}}
\medskip
\centerline{\it California Institute of Technology, Pasadena, CA 91125, USA}
\bigskip
\bigskip
\bigskip
\centerline{{\it Presented at the Workshop ``Frontiers in Quantum 
Field Theory''}}
\centerline{{\it in Honor of the 60th birthday of Keiji Kikkawa}}
\centerline{{\it Osaka, Japan \quad December 1995}}
\vskip 1.0 truein
\parindent=1 cm

\begin{abstract}
T duality expresses the equivalence of
a superstring theory compactified on a manifold $K$ to
another (possibly the same) superstring theory compactified on a dual manifold
$\tilde{K}$.  The volumes of $K$ and $\tilde{K}$ are inversely proportional.
In this talk we consider two M theory generalizations of T duality:
(i)  M theory compactified on a torus is 
equivalent to type IIB superstring theory
compactified on a circle and ~(ii) M theory compactified on a cylinder is
equivalent to $SO(32)$ superstring theory 
compactified on a circle. In both cases
the size of the circle is proportional to the $-3/4$ power of the area
of the dual manifold.
\end{abstract}
\vfil\eject
\setcounter{page}{1}
\section{Introduction}
\vglue0.4cm

It is a pleasure for me to speak on the occasion of the 60th birthday of my
good friend, Keiji Kikkawa.  I met Keiji during my first visit to the
Aspen Center for Physics in the summer of 1969.  This was shortly after the
appearance of the famous paper of Kikkawa, Sakita, and Virasoro,\cite{kikkawa}
which introduced the idea of regarding the dual resonance model $n$-point
functions as the tree approximation of a unitary quantum theory, rather than
just as interesting functions for use in phenomenology.  This seemed very novel
and exciting to me at the time.  In Aspen, Keiji and I collaborated in a study
of the Regge asymptotic behavior of the KSV loop amplitudes.\cite{kikkawaa}  I
found the experience so pleasant that I have returned to Aspen almost every
summer since then.

Many years later, after we had learned that the formulas of the earlier era
described strings and were better suited to unfication than to hadrons, Kikkawa
and Yamasaki discovered the ``T duality'' arising from compactification of
closed string theories on a circle.\cite{kikkawab}  This marked the beginning
of a new line of attack on string theory that has proved to be extremely
fruitful.\cite{giveon}  T
dualities are now recognized as a particular class of discrete dualities 
that are valid order-by-order in perturbation theory ({\it i.e.}, 
they are perturbative), though 
T duality is non-perturbative on the world sheet.  In my talk today
I would like to describe two simple extensions of T duality in the context of
``M theory.''  (M theory is a conjectured quantum theory in 
eleven dimensions
whose low-energy effective description is 
eleven-dimensional supergravity.)  An
interesting aspect of these extensions is that they incorporate some of the
more modern non-perturbative duality symmetries.

The basic results that I wish to describe are represented pictorially in Fig.
1.  This figure combines two well-known perturbative T dualities and
two non-perturbative 
identifications.  The perturbative T duality on the left side of the
figure is between type IIB superstring theory on a circle and type IIA
superstring theory on a circle of reciprocal radius.\cite{dine}  The
perturbative T duality on the right side of the figure relates the $E_8 \times
E_8$ heterotic string on a circle and the $SO(32)$ heterotic string on a circle
of reciprocal radius.\cite{narain}  One non-perturbative fact
is that the IIA theory
in ten dimensions actually has a circular 
eleventh dimension whose radius is
proportional to the 2/3 power of the string coupling 
constant.\cite{townsend,witten} 
Another is that the $E_8 \times E_8$ heterotic
theory in ten dimensions actually has an 
eleventh dimension that is 
a line segment I
(or, equivalently, a $\ZZ_2$ orbifold of a circle).\cite{horava}  The length of
the line segment also scales as the 2/3 power of the string coupling constant.

What I propose to do in this 
lecture is to by-pass the IIA and $E_8 \times E_8$
theories and discuss the following two dualities:  1)  
the equivalence of M theory on a
two-torus $T^2$ and IIB superstring 
theory on a circle $S^1$; 2) the equivalence of M theory on a
cylinder $C$ and $SO(32)$ 
superstring theory on a circle $S^1$.  In each case we will find
that the area of the manifold $T^2$ and $C$ (at fixed shape) 
scales as the $-4/3$
power of the size of the circle $S^1$.  The wrapping of all relevant $p$-branes
on the compact spaces will be considered.  By requiring a detailed matching
of $p$-branes in each pair of dual theories  we achieve numerous
consistency tests of the overall picture and deduce relations among various
$p$-brane tensions.  The discussion of M/IIB duality is a review of
results reported previously,\cite{schwarz} whereas the M/$SO(32)$ duality 
has not been presented previously.

\section{BPS - Saturated p-Branes}

A useful technique for obtaining non-perturbative information about superstring
theories and M theory is to identify their BPS saturated $p$-branes.
$p$-branes are $p$-dimensional objects that are characterized by a
``tension'' $T_p$, which is the mass per unit $p$-volume (and thus has
dimensions of $(mass)^{p+1}$), and a suitable ($p +
1$)-dimensional world-volume theory.  In theories with enough supersymmetry
(all supersymmetric theories in ten or eleven dimensions, in particular) the
tension of a $p$-brane carrying a suitable conserved charge has a strict lower
bound proportional to that charge.  When there is equality ($T = Q$), the
$p$-brane is said to be BPS saturated.  Such $p$-branes belong to ``short''
representations of the supersymmetry algebra.  So long as the supersymmetry is
not broken, the tension of such a $p$-brane cannot be changed by any quantum
correction -- perturbative or non-perturbative.  This generalizes the
well-known fact that the photon belongs to a ``short'' representation of the
Poincar\'e group and must remain massless so long as gauge invariance remains
unbroken.

BPS saturated $p$-branes of superstring theories or M theory can be
approximated by classical solutions of the corresponding effective supergravity
theory that preserve half of the supersymmetry.  Such solutions are not exact
superstring solutions, of course, but they do demonstrate the existence of
particular $p$-branes, give their tensions correctly, and exhibit other
qualitatively correct properties.  Generally these objects can be regarded as
extremal black $p$-branes, 
{\it i.e.}, extremal black holes $(p = 0)$, extremal black
strings $(p = 1)$, etc.  In solving the supergravity equations to obtain
$p$-brane configurations it is sometimes necessary to include a source to match
a delta function singularity in the equations.  When the source is required,
one often speaks of a ``fundamental $p$-brane'' and when it isn't of a
``solitonic $p$-brane.''  It will not be necessary for us to keep track of this
distinction here, however.

The conserved charges carried by $p$-branes are associated with antisymmetric
tensor gauge fields
\begin{equation} 
A_n = A_{\mu_{1}...\mu_{n}} dx^{\mu_{1}} \wedge dx^{\mu_{2}} \wedge .... \wedge
dx^{\mu_{n}}.\end{equation} 
When these undergo gauge transformations $\delta A_n = d \Lambda_{n-1}$, the
field strength $F_{n+1} = dA_n$ is invariant.  A supergravity theory with such
a gauge field typically has two kinds of BPS-saturated $p$-brane solutions.
The electric $p$-brane has $p = n - 1$ and the dual magnetic $p$-brane has $p =
D - n - 3$.  When it is possible to reformulate the supergravity theory in
terms of a dual potential $\tilde{A}$ (satisfying $\tilde{F} = d \tilde{A} = *
F$ plus possible interaction corrections) there is no essential distinction
between ``electric'' and ``magnetic'' $p$-branes.  The two are interchanged in
the dual formulation.  In the case of an electric $p$-brane, the gauge field
couples to the $p$-brane world volume generalizing the well-known $j \cdot A$
interaction of charged point particles.  It is also worth noting that the
charges of dual electric and magnetic $p$-branes satisfy a generalized Dirac
quantization condition $Q_E Q_M \in 2\pi \ZZ$.\cite{nepomechie}

Let us now consider specific examples, beginning with type IIB supergravity
theory.  This theory contains a complex scalar field $\rho = \chi + ie^{-\phi}$,
and so the specification of a vacuum is characterized by the modulus
\begin{equation} 
\rho_0 = \langle\rho\rangle = {\theta_B\over 2\pi} + {i\over\lambda_B},
\end{equation} 
where $\lambda_B$ is the string coupling constant.  (Note that the analogous formula for $N=4$ theories in four dimensions has 
$\rho_0 = {\theta/ 2\pi} + {i/\lambda^2}$.) The $S$ duality 
group of type IIB
superstring theory is an $SL(2,\ZZ)$ under which $\rho$ transforms
nonlinearly in the usual way.  Therefore, $\rho_0$ may be restricted to the usual
fundamental region.  The IIB theory also has a pair of two-form potentials
$B_{\mu\nu}^{(I)}$ that transform as a doublet of the $SL(2,\ZZ)$ duality
group.  Therefore, the associated electric 
one-branes (strings) carry a pair of
$B$ charges.  Suitably normalized, they can be chosen to be a pair of
relatively prime integers ($q_1, q_2$).  Setting $\theta_B = 0$ (to keep
things simple), the tensions (in the IIB string metric) are~\cite{schwarz}
\begin{equation} 
T_1^{(B)}(q_{1}, q_{2}) = 
\left(q_1^2 \lambda_B + {q_2^2\over \lambda_B} \right)^{1/2}
T_1^{(B)}.  \label{notheta}
\end{equation} 
In the canonical metric $T_1^{(B)}$ is a constant, and thus $T_1^{(B)} \sim
\lambda_B^{-1/2}$ in the string metric. Note that in the string metric
only $T_1^{(B)}(1,0)$  is finite as $\lambda_B \rightarrow 0$.
Therefore it is natural to regard it as the fundamental string and the others
as solitons, though they are all mapped into one another by the duality group.
The dual
five-branes carry a pair of magnetic $B$ charges.  The IIB theory
also contains a 
four-form potential $A_4$ with a self-dual field strength $F_5$.
As a result, electric and magnetic charge are identified in this case and
carried by a self-dual three-brane.

The M theory story is somewhat simpler, since the only antisymmetric tensor
gauge field in 
eleven-dimensional supergravity is a three-form $A_3$.  There are
associated electric two-brane and magnetic 
five-brane solutions.  It is apparently
not possible to replace $A_3$ by a dual six-form potential, so the
electric--magnetic distinction is meaningful in this case.  (The two-brane is
``fundamental'' and the five-brane is ``solitonic'' in the sense described
earlier.)

In the case of type I or heterotic strings the relevant low-energy theory is
$N=1$ $D=10$ supergravity coupled to an $E_8 \times E_8$ or $SO(32)$ super
Yang--Mills multiplet.  In this case the relevant antisymmetric tensor is a
two-form $B_{\mu\nu}$.  The associated electric one-brane is the heterotic
string.   There is also a five-brane, which is the
magnetic dual of the heterotic string.
Type I strings do not carry a conserved charge, 
they are not BPS saturated, and therefore they can break.  
For these reasons, they can only be described reliably at weak
coupling when they are metastable.   Now that we have described the relevant
theories and their $p$-branes we can turn to the duality analysis.

\section{M/IIB Duality}

The duality described by the left side of Figure 1 relates  M theory
compactified on a two-torus and IIB superstring theory compactified on a
circle.  The torus is described by its area $A_M$ (in the canonical
eleven-dimensional metric) and a modular parameter 
$\tau$, which may be taken to
lie in the fundamental region of the $SL(2,\ZZ)$ modular group.  The
corresponding parameters of the IIB theory are the modulus $\rho_0$ and the
circumference of the circle $L_B$ (in the canonical 
ten-dimensional metric, which is the one that is invariant under $SL(2,\ZZ)$
transformations).  To
test the equivalence of these two constructions, we will examine the matching
of all $p$-branes in nine dimensions.  These include various wrappings of the
ones identified in the preceding section as well as new $p$-branes that arise
by Kaluza--Klein mechanisms.  The possible $p$-brane wrappings are depicted in
Figure 2.  As 
the figure shows, $p$-branes, for any $p$ between 0 and 5, can be
obtained by suitable wrapping of a $p^{\prime}$-brane
in either M theory or IIB theory.  The
identifications are straightforward for $p = 1,2,3,4$.  They also work for $p =
0,5$, but one must be careful to take account of Kaluza--Klein effects in these
cases.  (The 
Kaluza--Klein vector fields in nine dimensions -- arising due to isometries --
support additional electric 
zero-branes and magnetic five-branes.)  When the matching
is done correctly, one finds a one-to-one correspondence 
of $p$-branes and their
tensions in nine dimensions.  The details have been worked out 
previously.\cite{schwarz} 
Here, I will simply state the results and discuss their implications.

The most important result, perhaps, is that one must make the
identification~\cite{schwarz,aspinwall}
\begin{equation} 
\rho_0 = \tau.\end{equation} 
Thus, the geometric $SL(2,\ZZ)$ modular group of the torus is identified with
the non-perturbative $S$-duality group of the IIB theory!  The canonical
metrics $g^{(M)}$ and $g^{(B)}$ are related (after the compactifications) by
\begin{equation} 
g^{(M)} = (A_M^{1/2} T_2^{(M)}/ T_1^{(B)} ) g^{(B)}.
\end{equation} 
Here, $T_2^{(M)}$ is the M theory two-brane tension.  Both $T_2^{(M)}$ and
$T_1^{(B)}$ (introduced earlier) are constants that define scales and can be
set to unity without loss of generality, though I will not do that.  In terms
of these constants, the compactification scales $A_M$ and $L_B$ are related by
\begin{equation} 
(T_1^{(B)} L_B^2)^{-1} = {1\over (2\pi)^2} T_2^{(M)} A_M^{3/2}.
\label{between}
\end{equation} 
Thus, $L_B \sim A_M^{-3/4}$.  This means that if one compactifies the IIB
theory on a circle and lets the size of the circle vanish, 
while holding $\rho_0$ fixed, one ends up with M
theory in eleven dimensions!  Conversely, if one compactifies the M theory on
a torus and lets the torus shrink to zero at fixed shape, one ends up with a
chiral theory -- IIB superstring theory -- in ten dimensions.

The matching of $p$-branes and their tensions also yields a number of relations
among the tensions.  Not only does one learn the relation between M and IIB
tensions in eq. (\ref{between}), but also relations among the tensions of the
M and IIB theories separately.  For the M theory, one learns that the
five-brane tension is proportional to the square of the two-brane tension
\begin{equation} 
T_5^{(M)} = {1\over 2\pi} (T_2^{(M)})^2.
\end{equation} 
This formula implies that the product of electric and magnetic charges is the
minimum value allowed by the quantization condition.\cite{duff}  
For the $p$-branes of the IIB theory one finds that all of their tensions
can be expressed in terms of $T_1^{(B)}$ and the moduli
\begin{equation} 
T_1^{(B)} (q_1, q_2) = \Delta_q^{1/2} T_1^{(B)} \label{withtheta}
\end{equation} 
\begin{equation} 
T_3^{(B)} = {1\over 2\pi} (T_1^{(B)})^2
\end{equation} 
\begin{equation} 
T_5^{(B)} (q_1, q_2) = {1\over (2\pi)^2} \Delta_q^{1/2} (T_1^{(B)})^3,
\end{equation} 
where
\begin{equation} 
\Delta_q = \left(q_1 - {q_2 \theta_B\over 2\pi} \right)^2 \lambda_B +
{q_2^2\over\lambda_B}.  \label{yestheta}
\end{equation} 
Equations (\ref{withtheta}) and (\ref{yestheta}) generalize 
eq. (\ref{notheta}) to include $\theta_B \not= 0$.  Note that
the tension of all RR $p$-branes $\sim 1/\lambda_B$ in the string metric,
as expected for D-branes.\cite{polchinskia}

A number of amusing things are taking place in the $p$-brane matchings that
gave these relations.  Let me just mention one of them.  In matching 
zero-branes
there is a duality between the M theory two-brane and the IIB strings
that works as follows:\cite{schwarz}  The Kaluza--Klein excitations of the
strings on a circle correspond to wrappings of the two-brane on the torus.
Conversely, the wrappings of the $SL(2,\ZZ)$ family of strings on the circle
correspond to the Kaluza--Klein excitations of the membrane on the torus.

Let us pause for a moment to consider the physical meaning of what we have
shown.  The claim is that M theory on $T^2$ is the same thing as IIB theory on
$S^1$.  If one considers the common nine-dimensional theory, one might imagine
asking the question ``How many compact dimensions are there?''  This question
has two correct answers -- one and two -- depending on whether one thinks of M
theory or IIB theory.  This paradoxical situation has a simple resolution.
Fields that describe ``matter'' in one picture can describe ``metric'' in the
other and vice versa.  This situation already occurs for more conventional T
duality -- say, for the heterotic string on a torus.  In that case the duality
mixes up internal components of the metric with those of the 
two-forms and $U(1)$
gauge fields.  What is new in the present case is that ~(i) the dual compact
spaces have different dimensions, ~(ii) certain components of the IIB theory
metric correspond to components of the 
three-form gauge field in M theory, ~(iii)
certain components of the M theory metric correspond to the complex scalar
$\rho$ of the IIB theory.  Despite these differences of detail, the basic
concept is the same.  Another important distinction, of course, is that the
generalization of T duality considered here encodes non-perturbative features
of the theory.

\section{M/SO(32) Duality}

Let us now consider the duality depicted in the right-hand portion of Figure 1.
This is the equivalence of M theory compactified on a cylinder with $SO(32)$
superstring theory compactified on a circle.  
Note that we do not specify whether the
$SO(32)$ theory is type I or heterotic.  The reason, of course, is that they
are different descriptions of a single theory,\cite{witten,polchinski} so it
is both of them.  
The dilaton in one description is the negative of the dilaton in the
other, and so the coupling constants are related by $\lambda_H^{(O)} =
(\lambda_I^{(O)})^{-1}$, where $O$ represents $SO(32)$, $H$ represents
heterotic, and $I$ represents type I.

Recall that the $SO(32)$ theory has two BPS saturated $p$-branes: the heterotic
string and its magnetic dual, which is a five-brane.  As discussed earlier, the
type I string is not BPS saturated and will not be considered in our analysis.
The $p$-branes arising from compactification on a circle are straightforward
to work out and are depicted in Figure 3.  As in Section 3, one should also
be careful to include $p$-branes of Kaluza--Klein origin.

M theory, before compactification, has a two-brane and a five-brane, but new
issues arise when one considers compactification on a manifold with boundaries,
and so we must first get that straight.  Consider first the Horava--Witten
picture~\cite{horava} -- M theory  on $\RR^{10} \times I$, an 
eleven-dimensional space-time with two parallel ten-dimensional 
boundaries. This is the non-perturbative description of the $E_8 \times
E_8$ heterotic string, just as
M theory on $\RR^{10} \times S^1$ describes the type
IIA superstring.  We know that $E_8 \times E_8$ theory, regarded as
ten-dimensional has a one-brane (the heterotic string) 
and a dual five-brane.  So we
must ask what these look like in eleven dimensions.  There is only one sensible
possibility.  An $E_8 \times E_8$ heterotic ``string'' is really a cylindrical
two-brane with one boundary attached to each boundary of the space time.  Thus,
an $E_8 \times E_8$ heterotic string can be viewed as a ribbon with one $E_8$
gauge group living on each boundary.  This seems to be the only allowed
two-brane configuration, since any other would give rise to a two-brane that
remains two-dimensional in the weak coupling limit in which the space-time
boundaries approach one another.  For the five-brane, the story is just the
reverse.  It must not terminate on the space-time boundaries, but can exist as
a closed surface in the bulk.  This is required so that it can give a
five-brane and not a four-brane in the weak coupling limit.  Subsequent
reduction on a circle to nine dimensions gives the wrapping possibilities
depicted in Figure 3.

The cylinder $C$ has a height $L_1$ and a circumference $L_2$.  These are
convenently combined to give an area $A_C = L_1L_2$ and a shape ratio $\sigma =
L_1/L_2$.  The circumference of the circle for the $SO(32)$ theory
compactification is denoted $L_O$
(in the heterotic string metric).  Also, the $p$-brane tensions of the
$SO(32)$ theory are denoted $T_1^{(O)}$ and $T_5^{(O)}$.
Now, guided by Figure 3, we again match $p$-branes in nine dimensions, just as
we did in the previous section.  One finds that the analog of 
the identification $\rho_0 = \tau$ is 
\begin{equation} 
\sigma = \lambda_H^{(O)} = (\lambda_I^{(O)})^{-1}  = L_1/L_2.
\label{sigmaeqn}
\end{equation} 
This means that when the spatial cylinder is a thin ribbon $(\sigma \ll 1)$ the
heterotic string is weakly coupled, whereas when it is a thin tube $(\sigma \gg
1)$, the type I string is weakly coupled.  
The analog of eq. (\ref{between}) is
\begin{equation} 
(T_1^{(O)} L_O^2)^{-1} = {1\over (2\pi)^2} T_2^{(M)} A_C^{3/2} \sigma^{-1/2}.
\end{equation} 
Aside from the factor of $\sigma^{-1/2}$ the equations look the same.  As
before, for fixed $\sigma, L_O \sim A_C^{-3/4}$.  Thus, the
$SO(32)$ theory in ten dimensions can be obtained by shrinking the 
cylinder to a point.

As in Section 3, the $p$-brane matching in nine dimensions gives various
tension relations.  The only new one is
\begin{equation} 
T_5^{(O)} = {1\over (2\pi)^2} \left({L_2\over L_1}\right)^2 (T_1^{(O)})^3.
\label{t5eqn}
\end{equation} 
Combining eqs. (\ref{sigmaeqn}) and (\ref{t5eqn}),
one learns that in the heterotic string metric, where
$T_1^{(O)}$ is constant, $T_5^{(O)} \sim (\lambda_H^{(O)})^{-2}$,
as expected for a soliton.  On the other
hand, in the type I string metric $T_1^{(O)} \sim {1/\lambda_I^{(O)}}$ and
$T_5^{(O)} \sim {1/\lambda_I^{(O)}}$, as expected for
$D$-branes.\cite{polchinskia}

In the case of M/IIB duality we found that the $SL(2,\ZZ)$ modular group of
the torus corresponded to the $S$-duality group of the IIB theory.  In the
present case of M/$SO(32)$ duality, the cylinder does not have an analogous
modular group.  However, the interchange $L_1 \leftrightarrow L_2$ 
(or $\lambda_I \leftrightarrow \lambda_H$) 
corresponds to the strong/weak duality transformation of the
$SO(32)$ theory that relates the perturbative heterotic limit to the
perturbative type I limit.

\section{Conclusion}

On several occasions Keiji Kikkawa has pioneered concepts that have led to
important advances in string theory.
The lessons of T duality are still being learned.  In particular, 
the generalization of T
duality to M theory described here exhibits the non-perturbative equivalence
of all known superstring theories, realizing a dream I have had for many years.
Other generalizations 
of T duality have been found,\cite{sen} and there may still be more
to come.  We owe Keiji a debt of gratitude for pointing us in this direction.

\vfill\eject

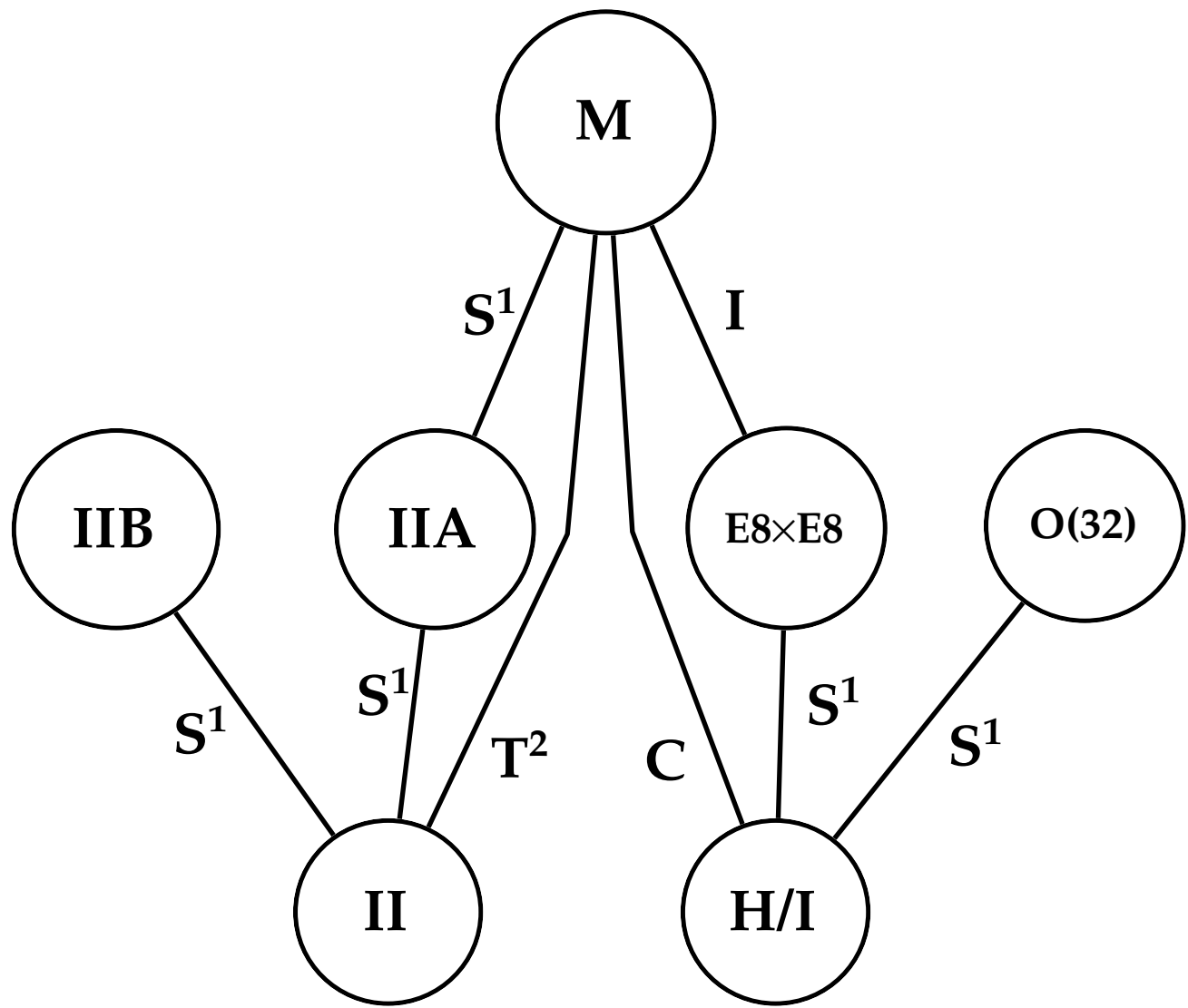
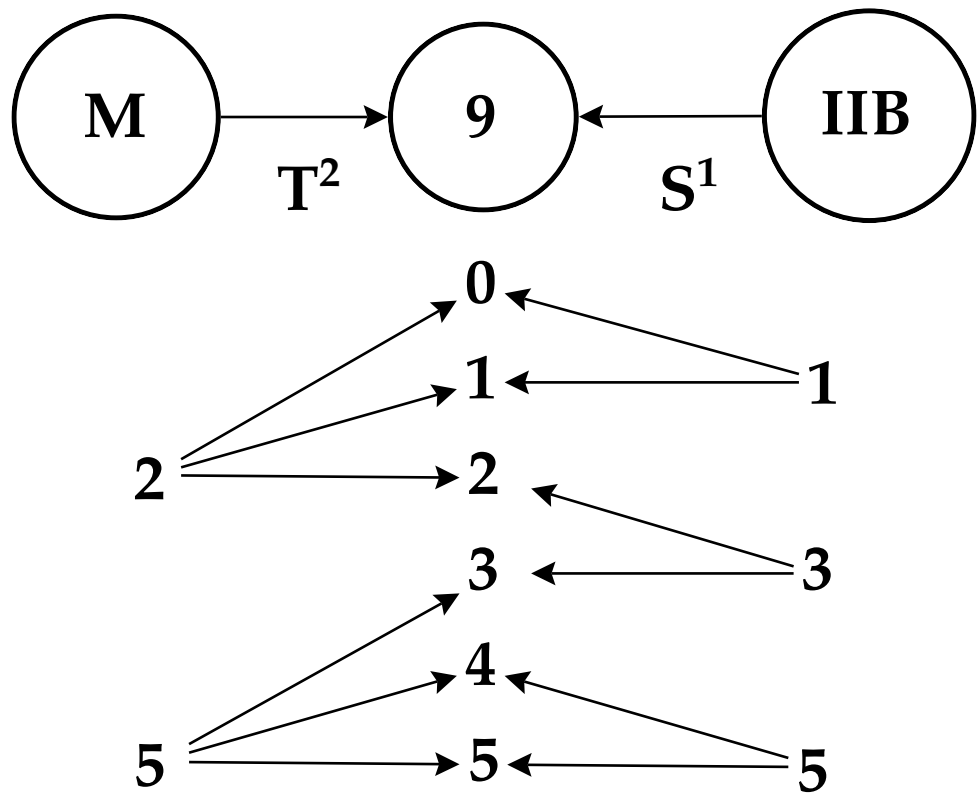
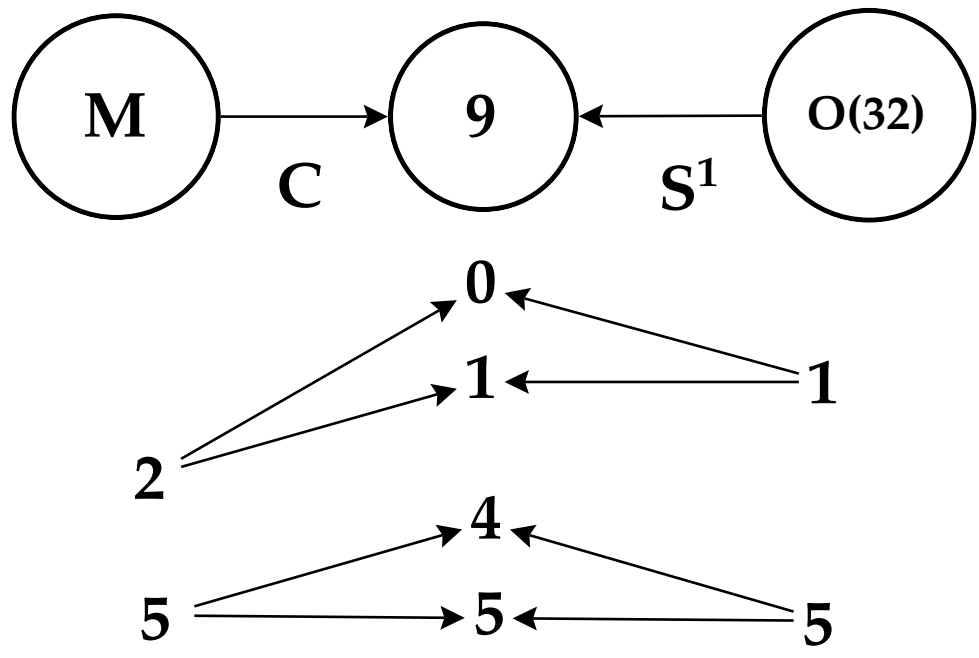
\end{document}